# New Reversal Mode in Exchange Coupled Antiferromagnetic/Ferromagnetic Disks: Distorted Viscous Vortex

Dustin A. Gilbert,[a] Li Ye,[a] Aïda Varea,[b] Sebastià Agramunt-Puig,[c] Nuria del Valle,[c] Carles Navau,[c] José Francisco López-Barbera,[c,d] Kristen S. Buchanan,[e] Axel Hoffmann,[f] Alvar Sánchez,[c] Jordi Sort,[c,g] Kai Liu,[a,*] and Josep Nogués[c,d,g,*]

[a]Physics Department, University of California, Davis, CA, USA

[b]MinD-in2UB, Electronics Department, Universitat de Barcelona, Martí i Franquès 1, 08028, Barcelona, Spain

[c]Departament de Física, Universitat Autònoma de Barcelona, 08193 Bellaterra (Barcelona), Spain

[d]ICN2 – Institut Catala de Nanociencia i Nanotecnologia, Campus UAB, 08193 Bellaterra (Barcelona), Spain

[e]Department of Physics, Colorado State University, Fort Collins, CO, USA

[f]Materials Science Division, Argonne National Laboratory, Argonne, IL, USA

[g]Institució Catalana de Recerca i Estudis Avançats (ICREA), Barcelona, Spain

* Corresponding authors: kailiu@ucdavis.edu; Josep.Nogues@uab.cat.

**Abstract**

Magnetic vortices have generated intense interest in recent years due to their unique reversal mechanisms, fascinating topological properties, and exciting potential applications. Additionally, the exchange coupling of magnetic vortices to antiferromagnets has also been shown to lead to a range of novel phenomena and functionalities. Here we report a new magnetization reversal mode of magnetic vortices in exchange coupled $Ir_{20}Mn_{80}/Fe_{20}Ni_{80}$ microdots: distorted viscous vortex reversal. Contrary to the previously known or proposed reversal modes, the vortex is distorted close to the interface and viscously dragged due to the uncompensated spins of a thin antiferromagnet, which leads to unexpected asymmetries in the annihilation and nucleation fields. These results provide a deeper understanding of the physics of exchange coupled vortices and may also have important implications for applications involving exchange coupled nanostructures.

Magnetic vortices have long been studied and remain a topic of current interest due to their fascinating fundamental properties and topological characteristics.[1-4] This magnetization state, which arises from the competition between magnetostatic, exchange and anisotropy energies in nanostructured ferromagnet (FM) materials, is characterized by a (counter-)clockwise in-plane curl of the magnetization (chirality) around an up or down out-of-plane central core (polarity).[5,6] Recent demonstrations of chirality and polarity control[7-11] have triggered renewed interests in these entities for spintronic applications,[12-14] artificial Skyrmion lattices,[15] and even biomedical applications.[16]

On the other hand, exchange bias [i.e., nominally the exchange coupling between a FM and an antiferromagnet (AF)] has received ever increasing interests across many emerging frontiers of condensed matter physics, e.g. , multiferroics,[17,18] chiral ordering and exchange bias induced by Dzyaloshinskii-Moriya interactions,[19,20] control of quantum magnets,[21] AF spintronics,[22,23] and triplet pairing in superconducting exchange biased heterostructures.[24] When a magnetic vortex is coupled to an AF (exchange bias) novel effects emerge, e.g., biased vortex reversal hysteresis loops, reversible non-zero remnant magnetization states, tunable angular dependent reversal modes, chirality-control, adjustable magnetization dynamics, or suppressed stochastic effects.[7, 25-33] These effects, which occur due to the imprinting of different magnetic states in the AF[34,35] can lead to additional functionalities in vortex structures. Furthermore, it has been predicted that in exchange coupled structures the vortex cores may be *tilted* along their thickness (in contrast with conventional vortices where the cores are straight) due to the pinning effects of the AF.[36,37] Such a tilted structure leads to additional asymmetries in the hysteresis loops, which are correlated with structural and magnetic parameters (e.g., dot geometry or AF/FM exchange strength). In fact, vortices exchange coupled to AFs have been prominently featured in key technologies such as magnetic random access memories and sensors.[38]

In this article we report a viscous vortex reversal mode in AF/FM exchange biased dots with a thick FM layer and varying AF thicknesses. By changing the AF thickness, $t_{AF}$, its anisotropy energy is systematically tuned, thus changing the rigidity of the AF spin structure from weak ("draggable" by the FM layer) to rigid. This leads to a viscous vortex reversal mechanism in the dots, which deviates from the standard, biased and tilted vortex reversals.

**Results**

**Major hysteresis loops**

Major hysteresis loops of $Fe_{20}Ni_{80}(30nm)/Ir_{20}Mn_{80}(t_{AF})$ (FeNi/IrMn) films are shown in Fig. 1(a) and the $H_C$ and $H_E$ trends in Fig. 1(b). These plots show that the $t_{AF}=0$ film has a very small coercivity ($H_C$=0.4 Oe) and no bias ($H_E$=0), while the $t_{AF}$= 3 nm film has a significantly increased $H_C$ (28 Oe) and a small $H_E$ (3.8 Oe). For $t_{AF}$>3 nm $H_C$ decreases (2.9–4.3 Oe), but $H_E$ is established (56–81 Oe). This behavior has been previously attributed to the anisotropy of the AF. Specifically, for a thin AF the anisotropy is exceedingly weak and is viscously dragged by the FM while it reverses,[39] leading to enhanced $H_C$, but no bias. For the thicker AF the anisotropy is sufficiently large so that the spins remain rigidly oriented after the field cooling process.[40, 41]

Major hysteresis loops for dots with 1μm and 1.5μm diameter are shown in Figs. 1(c-f). The symmetric pinched shape of the loops without AF, Fig. 1(c), is characteristics of a vortex state reversal. For dots with a thin IrMn layer ($t_{AF}$=3 nm), the major loops, Fig. 1(d), exhibit a much larger coercivity and a pronounced asymmetry. However, these dots do not show appreciable exchange bias, indicating that the AF has weak anisotropy and is dragged during the FM reversal. For dots with thicker ($t_{AF} \geq 5$ nm) IrMn layers, Figs. 1(e, f), the exchange bias is clearly established, and the $H_C$ is less than that of the $t_{AF}=3$ nm samples. Close inspection of the loops reveals asymmetries in their shape, particularly for $t_{AF}=5$ nm, suggesting the presence of locally pinned spins. Interestingly, the major loop shape for dots with an AF is quite complex and not indicative of any traditional magnetization reversal modes.

A possible reversal mechanism proposed previously in exchange biased dots is the "tilted-vortex" model.[36, 37] In this model the interfacial moments in the FM are pinned by the exchange coupling to the AF and the vortex core position at this interface is displaced from the center, while further away from the interface the vortex is more centered, hence the core is tilted. As shown schematically in Fig. 2(a), the reversal mechanism is reflected in the major loops by the asymmetry of the positive and negative annihilation fields, $H_A^+$ and $H_A^-$ respectively, after offsetting the $H_E$. That is, in an unbiased vortex $H_A^+=-H_A^-$, while in a biased vortex $H_A^+-H_E=-(H_A^- -H_E)$, and in a tilted vortex $H_A^+ -H_E \neq -(H_A^- -H_E)$. Furthermore, the nucleation field, $H_N$, should always be equally biased; for normal, biased and tilted vortex reversal $H_N^+- H_E=-(H_N^- -H_E)$. We can thus define variables $\Delta H_A= H_A^+ + H_A^- -2H_E$ and $\Delta H_N= H_N^+ + H_N^- -2H_E$. By identifying $\Delta H_A$ and $\Delta H_N$ these three reversal behaviors can be uniquely identified: $\Delta H_A=\Delta H_N=H_E=0$ for unbiased vortices, $\Delta H_A=\Delta H_N=0$ and $H_E\neq 0$ for biased vortices, and $\Delta H_A\neq 0$, $\Delta H_N=0$ and $H_E\neq 0$ for tilted vortices. The trends for $\Delta H_A$ and $\Delta H_N$ are shown in Figs. 2(b,c), where the nucleation/annihilation fields are determined from intercepts of the linear extrapolations of magnetization before and after nucleation/annihilation. It can be seen that indeed there is an asymmetry in $H_A$, suggesting a tilted-vortex reversal. However, there is also an asymmetry in the nucleation field, which is unexpected in any of the reversal behaviors discussed above.

While the analytical theory for exchange bias induced vortex tilting by Guslienko and Hoffmann can qualitatively explain several of the experimentally observed effects,[36, 42] some of the main experimental observations cannot be accounted for. For example (i) the experimental $\Delta H_A$ is not proportional to the macroscopic $H_E$ as assumed in the model; (ii) decreasing the dot diameter, $\Delta H_A$ decreases rather than increases as obtained from the calculations; or (iii) there is a $H_N$ asymmetry, which the theory assumes to be absent. These discrepancies suggest that the microscopic magnetic structure is far more complex than the one assumed in the theory. For instance, the model is mainly based on the depth dependence of the effective exchange bias field, but it neglects the non-uniform spin structure at the interface. In particular, it is well accepted that the exchange bias effect can be related to pinned and unpinned uncompensated spins in the AF/FM interface,[43-48] giving rise to loop shifts and coercivity enhancements, respectively. In the biased vortex case, we can naively assume that the pinned uncompensated spins will be parallel to the cooling field, while the unpinned ones will form a curl mimicking the FM vortex. Hence, while the vortex near the FM/AF interface experiences a complex energy landscape, away from this interface it behaves more like a conventional vortex. These additional interface effects, which should be enhanced for smaller sizes, could give rise to the discrepancies between the theory and experiments.

**Micromagnetic simulations**

To highlight the origin of the novel reversal mode micromagnetic simulations were conducted. The simulated loop, shown in Fig. 3(a), exhibits a pinched loop shape, typical of vortex reversal, shifted along the field axis (with $H_E$=128 Oe and $H_C$=54 Oe), in good qualitative agreement with the experimental results. The increasing and decreasing field branches of the loop are shown to have different $H_A$ and $H_N$ [see Fig. 3(b)], with $\Delta H_A$ and $\Delta H_N$ of about 8 Oe, reproducing the asymmetries observed experimentally. To elucidate the origin of this asymmetry we examine the spin maps of each layer. In Fig. 3(c) we plot the orientation of the spin moments (at H~$H_C$) at the AF/FM interface and the top FM surface of the dot [labeled layers 5 and 1 in Fig. 3(d)] in green and black, respectively. The color contrast in Fig. 3(c) identifies the magnetization difference in the two layers ($m_{y1}$-$m_{y5}$) as red (positive) and blue (negative). The first remarkable result is that the core of the first and fifth layers seems to be at the same position (within one micromagnetic cell, 6 nm), indicating no vortex tilt, in contrast with theoretical predictions.[36, 37] However, as demonstrated in Fig. 3(c) the vortices exhibit a clear distortion. While layer 1 (green) has a near perfect vortex structure, the interfacial spins in layer 5 tend to tilt towards the FC direction [see Figs. 3(d,e)]. The distortion is more pronounced along the ascending-field branch [Fig. 3(c) right panels] compared to the descending-field branch [left panels], as illustrated by the more intense background color. The origin of the major loop asymmetries seems to be related to different degree of distortion of the vortex structure close to the AF. This variation in the interfacial coupling is manifested differently in $H_A$ and $H_N$ depending on the previous saturation states and the field cooling direction.

**First order reversal curve (FORC) analysis**

To gain a detailed understanding of the dot magnetization reversal, we have performed FORC studies on the microdot arrays [see Electronic Supplementary Information (ESI)].[49-53] The FORC diagrams for the unbiased dots [Figs. 4(a, b)] show "butterfly"-like features of a standard vortex reversal with three main peaks, identified in Fig. 4(a) and discussed in the ESI (Fig. S1): peak *i* corresponds to the initial vortex nucleation from positive saturation and subsequent annihilation approaching positive saturation;[51] peak *ii* corresponds to re-nucleation from negative saturation, and is accompanied by a negative region that reflects the slope change along successive FORCs;[52] peak *iii* identifies subsequent annihilation to positive saturation, manifesting asymmetries in the dot shape.[50] For $t_{AF}=0$, features *i* and *ii* are of similar intensity [Figs. 4(a, b)] since the nucleation events are symmetric under field inversion.

For $t_{AF}=3$nm, intensities of the FORC features *i* and *ii* become asymmetric [Figs. 4(c, d)], indicating a deviation from the conventional vortex reversal and asymmetric magnetization reversal processes. The asymmetry is even more pronounced for $t_{AF}=5$nm, where feature *i* has largely vanished. The suppression of feature *i* indicates a much-reduced irreversibility associated with the vortex nucleation/annihilation near the positive saturation, while the enhanced FORC peak *ii* shows that the primary irreversibility is due to the vortex nucleation/annihilation near the negative saturation. This vortex reversal asymmetry is consistent with a depth-dependent magnetization configuration in the dots,[51, 54] since the pinning induced by the AF is stronger at the FM/AF interface than at the FM free surface, as suggested by the simulations. In both $t_{AF}=3$nm and 5nm peak *ii* shifts to a larger *local* coercivity ($H_C$*- see ESI), consistent with the proposed viscous drag reversal. For $t_{AF}=5$nm the entire FORC distribution is shifted towards negative $H_B$ as the exchange bias is established [Figs. 4(e, f)].

Finally, for $t_{AF}=7$nm, shown in Figs. 4(g) and 4(h), (and 9nm, not shown) the FORC distribution returns to a "butterfly"-like feature set. In addition, the nucleation/annihilation features are of comparable intensity. Recalling that $\Delta H_A$ is nearly zero for these samples [Fig. 2(a)], this indicates that the reversal involves simply biased vortices, *not* tilted vortices.

**Discussions**

The reversal mechanism observed for $t_{AF}$=3 and 5 nm deviates from the three established behaviors discussed earlier (vortex, shifted vortex and tilted vortex), none of which predict an asymmetry in $H_N$. Originating from the drag of the AF and accompanied by a distortion of the vortex structure rather than a tilt (as shown by simulations), this magnetization reversal mechanism may be viewed as distorted viscous vortex reversal. Remarkably, the dependence of $H_E$ and $H_C$, $\Delta H_A$ and $\Delta H_N$ and the evolution of the FORC features on $t_{AF}$ seem to indicate that the new reversal mode is dominated by the unpinned uncompensated spins, which explains its differences from the proposed tilted-vortex mode. Nevertheless, the asymmetries related to this new mechanism should be enhanced for thicker FM layers (where the vortex distortion should increase) and moderately thin AFs (where the AF has weaker anisotropy and the drag should be larger). Even in nanostructures with thick AFs, distorted viscous vortex reversal may still emerge if the temperature is sufficiently increased so that the AF anisotropy is concomitantly weakened.[55] Interestingly, although the thickness of the FM layers in AF/FM dots is typically on the 10 nm scale, some hints of reversal asymmetries probably linked to this new reversal mode can be found in the literature.[4, 30, 55, 56] Note that the viscous drag of the magnetization due to the AF also occurs in thin films. However, contrary to what is observed in nanostructures, in thin films the net effect of this viscous drag is merely an increase in coercivity without any changes in the magnetization reversal modes.[57, 58] Importantly, the dragging of the AF layer is not only a general feature of exchange biased dots, but may also be relevant for virtually *any* exchange coupled system,[59] e.g., in magnetically hard/soft exchange coupled nanostructures[60-62] where the harder layer has insufficient anisotropy to pin the softer layer.

The asymmetries inherent to the distorted viscous vortex reversal may have practical implications in the performance of magnetic devices based on exchange coupling. Thus, possible effects of the distorted viscous vortex reversal should be taken into account in the design of such devices (e.g., tuning the thickness of the AF or FM layers or operating temperature to avoid this effect).

**Conclusions**

In summary, we have found a new distorted viscous vortex reversal mode in exchange biased FeNi/IrMn dots with varying AF thicknesses. Unbiased dots reverse via a vortex state, while dots with an AF layer undergo a much more complex reversal process: dots with thin AF layers reverse via a distorted viscous vortex state with an enhanced coercivity; once the AF layer is thick enough to have sufficient anisotropy energy, the magnetization reverses via a biased vortex state, and the coercivity enhancement is suppressed. This viscous vortex reversal mode and the asymmetries in the annihilation and nucleation fields are beyond the current understanding of exchange coupled vortices, and offer interesting implications for device applications.

**Methods**

Arrays of circular nanodots with diameter of 0.5, 1.0 and 1.5 µm and vertical structure of Ta(5nm)/Fe$_{20}$Ni$_{80}$(30nm)/Ir$_{20}$Mn$_{80}$(t$_{AF}$)/Pt(2nm) [t$_{AF}$=0–9nm] were fabricated on a naturally oxidized Si(001) substrate by electron-beam lithography and DC magnetron sputtering from composite targets in 1.5mTorr Ar. The FM layer was kept deliberately thick to promote tilted vortex reversal.[36, 37] Arrays with a common AF thickness were fabricated in a single run, with the other arrays shadowed by a mask. The AF orientation was set by heating the sample to 520 K (above the blocking temperature of IrMn, $T_B$=420 K) then cooling to room temperature in an in-plane magnetic field, $H_{FC}$=2 kOe. Hysteresis loops and first-order reversal curve (FORC) measurements were recorded at room temperature using a longitudinal magneto-optical Kerr effect (MOKE) setup, following prior procedures,[49, 50, 63, 64] with loops measured along the cooling field axis, iteratively averaged at a rate of 7 Hz for ~1000 cycles.

Simulations were conducted using the same geometric constructions as the experimental system by iteratively solving Brown's static equations[65] using a 6 nm cubic mesh (consistent with the exchange length of FeNi[66]), making the simulated FM 5 cells thick. The polycrystalline FeNi was simulated using an exchange stiffness $A=1.3\times10^{-11}$ J/m, a saturation magnetization $M_S=8\times10^5$ A/m, and magnetocrystalline anisotropy $K_U=0$. The IrMn was modeled as 84% non-magnetic material and 16% (900 cells) randomly distributed magnetically contributing cells, representing uncompensated spins. The contributing cells are further divided into pinned and rotatable cells[43-48] in a ratio of 4:5, giving a moderate loop shift, $H_E$, and coercivity, $H_C$. The pinned cells have their magnetization ($M_S=8\times10^5$ A/m) fixed along the field-cool (FC) direction, while the unpinned ones have a large uniaxial anisotropy ($K_U=5\times10^5$ J/m) in the FC direction. These uncompensated spins interact via exchange (assuming $J_{AF-FM}=J_{FM-FM}$) and magnetostatic interactions with the FM spins but only magnetostatically among themselves (since they represent the equivalent of isolated uncompensated spins in experimental systems). Since the results depend on the spatial distribution of the pinned and unpinned spins, the presented results are the average of 8 different simulated configurations.

**Acknowledgements**

This work was supported by the US NSF (DMR-1008791 and ECCS-1232275), the 2014-SGR-1015 project of the Generalitat de Catalunya, and MAT2010-20616-C02, CSD2007-00041 and MAT2012-35370 projects of the Spanish Ministerio de Economía y Competitividad (MinECO). Work at Argonne was supported by the U. S. Department of Energy, Office of Science, Materials Science and Engineering Division. Fabrication was performed at the Center for Nanoscale Materials, which is supported by DOE, Office of Science, Basic Energy Science under Contract No. DE-AC02-06CH11357. KL acknowledges support from the NSFC (11328402). AS acknowledges a grant from the ICREA Academia, funded by the Generalitat de Catalunya. ICN2 acknowledges support from the Severo Ochoa Program (MinECO, Grant SEV-2013-0295).

**References**

1. J. I. Martin, J. Nogués, K. Liu, J. L. Vicent and I. K. Schuller, *J. Magn. Magn. Mater.*, 2003, **256**, 449-501.

2. K. Y. Guslienko, *J. Nanosci. Nanotech.*, 2008, **8**, 2745-2760.

3. R. P. Cowburn, D. K. Koltsov, A. O. Adeyeye, M. E. Welland and D. M. Tricker, *Phys. Rev. Lett.*, 1999, **83**, 1042-1045.

4. F. Liu and C. A. Ross, *J. Appl. Phys.*, 2014, **116**, 194307.

5. T. Shinjo, T. Okuno, R. Hassdorf, K. Shigeto and T. Ono, *Science*, 2000, **289**, 930-932.

6. A. Wachowiak, J. Wiebe, M. Bode, O. Pietzsch, M. Morgenstern and R. Wiesendanger, *Science*, 2002, **298**, 577-580.

7. W. Jung, F. Castaño and C. Ross, *Phys. Rev. Lett.*, 2006, **97**, 247209.

8. K. Yamada, S. Kasai, Y. Nakatani, K. Kobayashi, H. Kohno, A. Thiaville and T. Ono, *Nature Mater.*, 2007, **6**, 270-273.

9. M. Jaafar, R. Yanes, D. Perez de Lara, O. Chubykalo-Fesenko, A. Asenjo, E. M. Gonzalez, J. V. Anguita, M. Vazquez and J. L. Vicent, *Phys. Rev. B*, 2010, **81**, 054439.

10. R. K. Dumas, D. A. Gilbert, N. Eibagi and K. Liu, *Phys. Rev. B*, 2011, **83**, 060415.

11. M. Kammerer, M. Weigand, M. Curcic, M. Noske, M. Sproll, A. Vansteenkiste, B. Van Waeyenberge, H. Stoll, G. Woltersdorf, C. H. Back and G. Schuetz, *Nature Commun.*, 2011, **2**, 279.

12. L. Thomas, M. Hayashi, X. Jiang, R. Moriya, C. Rettner and S. S. P. Parkin, *Nature*, 2006, **443**, 197-200.

13. A. Ruotolo, V. Cros, B. Georges, A. Dussaux, J. Grollier, C. Deranlot, R. Guillemet, K. Bouzehouane, S. Fusil and A. Fert, *Nature Nanotechnol.*, 2009, **4**, 528-532.

14. V. S. Pribiag, I. N. Krivorotov, G. D. Fuchs, P. M. Braganca, O. Ozatay, J. C. Sankey, D. C. Ralph and R. A. Buhrman, *Nature Phys.*, 2007, **3**, 498-503.

15. L. Sun, R. X. Cao, B. F. Miao, Z. Feng, B. You, D. Wu, W. Zhang, A. Hu and H. F. Ding, *Phys. Rev. Lett.*, 2013, **110**, 167201.

16. D.-H. Kim, E. A. Rozhkova, I. V. Ulasov, S. D. Bader, T. Rajh, M. S. Lesniak and V. Novosad, *Nature Mater.*, 2009, **9**, 165-171.

17. Y. Tokunaga, Y. Taguchi, T. Arima and T. Tokura, *Phys. Rev. Lett.*, 2014, **112**, 037203.

18. W. Echtenkamp and C. Binek, *Phys. Rev. Lett.*, 2013, **111**, 187204.

19. J. Li, A. Tan, R. Ma, R. F. Yang, E. Arenholz, C. Hwang and Z. Q. Qiu, *Phys. Rev. Lett.*, 2014, **113**, 147207.

20. R. Yanes, J. Jackson, L. Udvardi, L. Szunyogh and U. Nowak, *Phys. Rev. Lett.*, 2013, **111**, 217202.

21. S. Yan, D.-J. Choi, J. A. J. Burgess, S. Rolf-Pissarczyk and S. Loth, *Nature Nanotechnol.*, 2015, **10**, 40.

22. I. Fina, X. Marti, D. Yi, J. Liu, J. H. Chu, C. Rayan-Serrao, S. Suresha, A. B. Shick, J. Zelezny, T. Jungwirth, J. Fontcuberta and R. Ramesh, *Nature Commun.*, 2014, **5**, 4671.

23. D. Ciudad, M. Gobbi, K. C. J., M. Eich, J. S. Moodera and L. E. Hueso, *Adv. Mater.*, 2014, **26**, 7561.

24. V. I. Zdravok, D. Lenk, R. Morari, A. Ullrich, G. Obermeier, C. Muller, H.-A. Krug von Niffa, A. S. Sidorenko, S. Horn, R. Tidecks and L. R. Tagirov, *Appl. Phys. Lett.*, 2013, **103**, 062604.

25. J. Mejía-López, P. Soto and D. Altbir, *Phys. Rev. B*, 2005, **71**, 104422.

26. J. Sort, A. Hoffmann, S. H. Chung, K. Buchanan, M. Grimsditch, M. Baró, B. Dieny and J. Nogués, *Phys. Rev. Lett.*, 2005, **95**, 067201.

27. Z.-P. Li, O. Petracic, J. Eisenmenger and I. K. Schuller, *Appl. Phys. Lett.*, 2005, **86**, 072501.

28. M. Tanase, A. Petford-Long, O. Heinonen, K. Buchanan, J. Sort and J. Nogués, *Phys. Rev. B*, 2009, **79**, 014436.

29. J. Sort, G. Salazar-Alvarez, M. D. Baró, B. Dieny, A. Hoffmann, V. Novosad and J. Nogués, *Appl. Phys. Lett.*, 2006, **88**, 042502.

30. J. Sort, K. Buchanan, V. Novosad, A. Hoffmann, G. Salazar-Alvarez, A. Bollero, M. Baró, B. Dieny and J. Nogués, *Phys. Rev. Lett.*, 2006, **97**, 067201.

31. O. G. Heinonen, D. K. Schreiber and A. K. Petford-Long, *Phys. Rev. B*, 2007, **76**, 144407.

32. K. S. Buchanan, A. Hoffmann, V. Novosad and S. D. Bader, *J. Appl. Phys.*, 2008, **103**, 07B102.

33. S. O. Parreiras, G. B. M. Fior, F. Garcia and M. D. Martins, *J. Appl. Phys.*, 2013, **114**, 203903.

34. G. Salazar-Alvarez, J. J. Kavich, J. Sort, A. Mugarza, S. Stepanow, A. Potenza, H. Marchetto, S. S. Dhesi, V. Baltz, B. Dieny, A. Weber, L. J. Heyderman, J. Nogués and P. Gambardella, *Appl. Phys. Lett.*, 2009, **95**, 012510.

35. J. Wu, D. Carlton, J. S. Park, Y. Meng, E. Arenholz, A. Doran, A. T. Young, A. Scholl, C. Hwang, H. W. Zhao, J. Bokor and Z. Q. Qiu, *Nature Phys.*, 2011, **7**, 303-306.

36. K. Y. Guslienko and A. Hoffmann, *Phys. Rev. Lett.*, 2006, **97**, 107203.

37. K. Y. Guslienko and A. Hoffmann, *J. Appl. Phys.*, 2007, **101**, 093901.

38. See e.g., T. Min, Y. Guo, and P. Wang, US Patent 7072208 B2 (2006); B. Dieny, US Patent Appl., 2015/0063019A1 (2015); P. Kasiraj and S. Maat, US Patent 7057862B2 (2004).

39. M. S. Lund, W. A. A. Macedo, K. Liu, J. Nogués, Schuller, I. K. and C. Leighton, *Phys. Rev. B*, 2002, **66**, 054422.

40. J. Nogués and I. K. Schuller, *J. Magn. Magn. Mater.*, 1999, **192**, 203-232.

41. M. Ali, C. Marrows, M. Al-Jawad, B. Hickey, A. Misra, U. Nowak and K. Usadel, *Phys. Rev. B*, 2003, **68**, 214420.

42. A. Hoffmann, J. Sort, K. S. Buchanan and J. Nogués, *IEEE Trans. Magn.*, 2008, **44**, 1968-1973.

43. J. Camarero, Y. Pennec, J. Vogel, S. Pizzini, M. Cartier, F. Fettar, F. Ernult, A. Tagliaferri, N. Brookes and B. Dieny, *Phys. Rev. B*, 2003, **67**, 020413.

44. S. Brück, G. Schütz, E. Goering, X. Ji and K. Krishnan, *Phys. Rev. Lett.*, 2008, **101**, 126402.

45. H. Ohldag, A. Scholl, F. Nolting, E. Arenholz, S. Maat, A. Young, M. Carey and J. Stöhr, *Phys. Rev. Lett.*, 2003, **91**, 017203.

46. R. L. Stamps, *J. Phys. D: Appl. Phys.*, 2000, **33**, R247-R268.

47. M. Fitzsimmons, B. Kirby, S. Roy, Z.-P. Li, I. Roshchin, S. Sinha and I. Schuller, *Phys. Rev. B*, 2007, **75**, 214412.

48. J. Nogués, S. Stepanow, A. Bollero, J. Sort, B. Dieny, F. Nolting and P. Gambardella, *Appl. Phys. Lett.*, 2009, **95**, 152515.

49. J. E. Davies, O. Hellwig, E. E. Fullerton, G. Denbeaux, J. B. Kortright and K. Liu, *Phys. Rev. B*, 2004, **70**, 224434.

50. R. K. Dumas, C. P. Li, I. V. Roshchin, I. K. Schuller and K. Liu, *Phys. Rev. B*, 2007, **75**, 134405.

51. R. K. Dumas, C.-P. Li, I. V. Roshchin, I. K. Schuller and K. Liu, *Phys. Rev. B*, 2012, **86**, 144410.

52. D. A. Gilbert, G. T. Zimanyi, R. K. Dumas, M. Winklhofer, A. Gomez, N. Eibagi, J. L. Vicent and K. Liu, *Sci. Rep.*, 2014, **4**, 4204.

53. D. A. Gilbert, J. W. Liao, L. W. Wang, J. W. Lau, T. J. Klemmer, J. U. Thiele, C. H. Lai and K. Liu, *APL Mater.*, 2014, **2**, 086106.

54. Z. P. Li, O. Petracic, R. Morales, J. Olamit, X. Batlle, K. Liu and I. K. Schuller, *Phys. Rev. Lett.*, 2006, **96**, 217205.

55. S. Thomas, D. Nissen and M. Albrecht, *Appl. Phys. Lett.*, 2014, **105**, 022405.

56. A. L. Dantas, G. O. G. Rebouças and A. S. Carriço, *IEEE Trans. Magn.*, 2010, **46**, 2311-2313.

57. J. Nogués and I. K. Schuller, *J. Magn. Magn. Mater.*, 1999, **192**, 203-232.

58. J. Nogues, J. Sort, V. Langlais, V. Skumryev, S. Surinach, J. S. Munoz and M. D. Baro, *Phys. Rep.*, 2005, **422**, 65.

59. J. E. Davies, O. Hellwig, E. E. Fullerton, J. S. Jiang, S. D. Bader, G. T. Zimanyi and K. Liu, *Appl. Phys. Lett.*, 2005, **86**, 262503.

60. Q. L. Ma, S. Mizukami, T. Kubota, X. M. Zhang, Y. Ando and T. Miyazaki, *Phys. Rev. Lett.*, 2014, **112**, 157202.

61. L. S. Huang, J. F. Hu, B. Y. Zong, S. W. Zeng, Ariando and J. S. Chen, *J. Phys. D: Appl. Phys.*, 2014, **47**, 245001.

62. D. Goll and A. Breitling, *Appl. Phys. Lett.*, 2009, **94**, 052502.

63. C. Pike, C. Ross, R. Scalettar and G. Zimanyi, *Phys. Rev. B*, 2005, **71**, 134407.

64. R. K. Dumas, K. Liu, C. P. Li, I. V. Roshchin and I. K. Schuller, *Appl. Phys. Lett.*, 2007, **91**, 202501.

65. W. F. Brown Jr., *Magnetostatic Principles in Ferromagnetism*, Amsterdam, North-Holland, 1962.

66. S. Agramunt-Puig, N. del Valle, C. Navau and A. Sanchez, *Appl. Phys. Lett.*, 2014, **104**, 012407.

**Figure Captions**

**Figure 1.** (Color online) (a) Hysteresis loops for continuous *films* of FeNi/IrMn with different $t_{AF}$. (b) Dependence of $|H_E|$ and $H_C$ on $t_{AF}$ for the films and 1.0 and 1.5 μm circular dots. Major hysteresis loops for the 1.0 and 1.5 μm *circular dots* with $t_{AF}$ of (c) 0 nm, (d) 3 nm, (e) 5 nm, and (f) 7 nm.

**Figure 2.** (Color online) Schematic hysteresis loops are shown in (a) for unbiased, biased, tilted, and viscous vortex reversals. The unbiased vortex reversal is shown in dotted grey for reference. The viscous vortex reversal shows little to no exchange field, but asymmetries in both the nucleation and annihilation. Symbols $H_N^+$ ($H_N^-$) and $H_A^+$ ($H_A^-$) represent nucleation field from positive (negative) saturation and annihilation to positive (negative) saturation, respectively. Measured dependence of (b) $\Delta H_A$ and (c) $\Delta H_N$ on $t_{AF}$ for d=1.0 and 1.5 μm circular dots. Error bars [smaller than symbol size in (b)] are determined by the radius of curvature of the measured data at the nucleation (annihilation) corner.

**Figure 3.** (Color online) (a) Simulated hysteresis loop of a FM disk with 504nm diameter and 30nm thickness pinned to an AF layer. (b) Composition of the top half of the loops (i.e., M>0) (black symbols) and the inverted bottom part (M<0, corrected for the loop shift (red symbols). (c) Spin maps of the top (black) and bottom (green) FM layers at H~$H_C$ for decreasing (left) and increasing (right) fields. The background intensity corresponds to the difference ($m_{y1}$-$m_{y5}$) as red (positive) and blue (negative), as discussed in the text. The bottom images are enlarged views of the highlighted areas. The cooling field ($H_{FC}$) and applied field ($H_{Appl}$) directions are shown by arrows. (d) Side- and (e) top-view schematic illustrations of the magnetic spins in a distorted vortex structure based on the simulations.

**Figure 4.** (Color online) FORC distributions for (left) 1.0 μm and (right) 1.5 μm diameter exchange biased dots with $t_{AF}$ of (a, b) 0 nm, (c, d) 3 nm, (e, f) 5 nm, and (g, h) 7 nm.

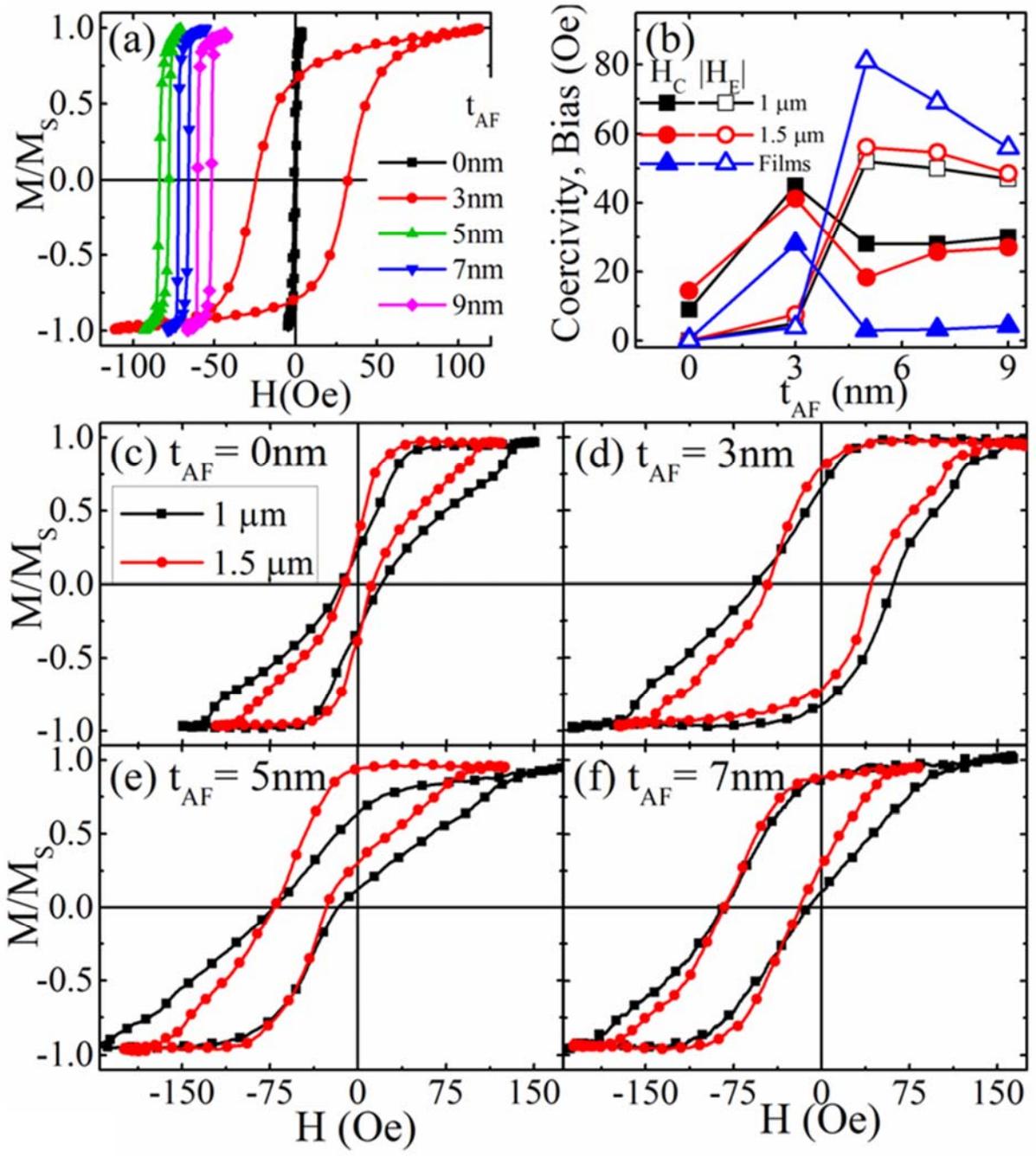

**Figure 1**

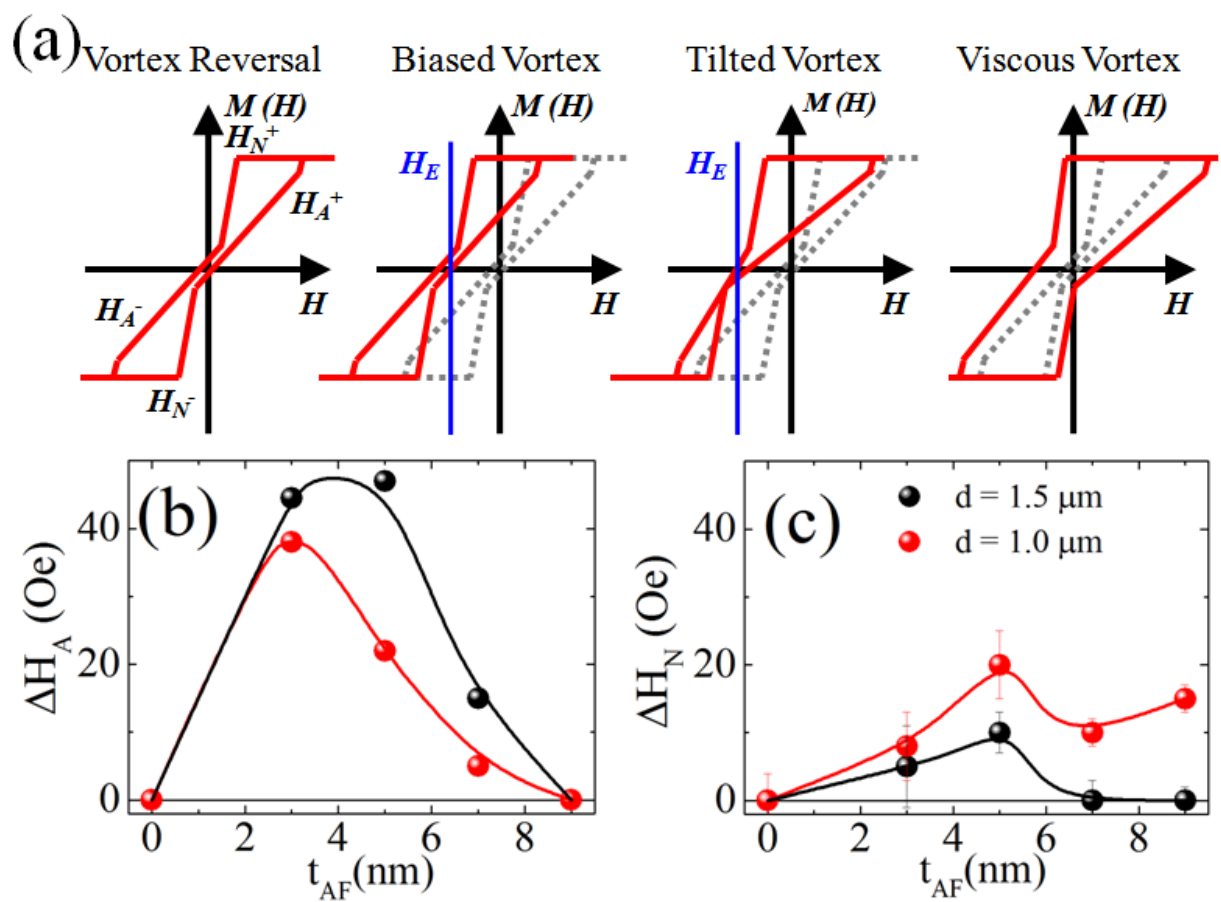

**Figure 2**

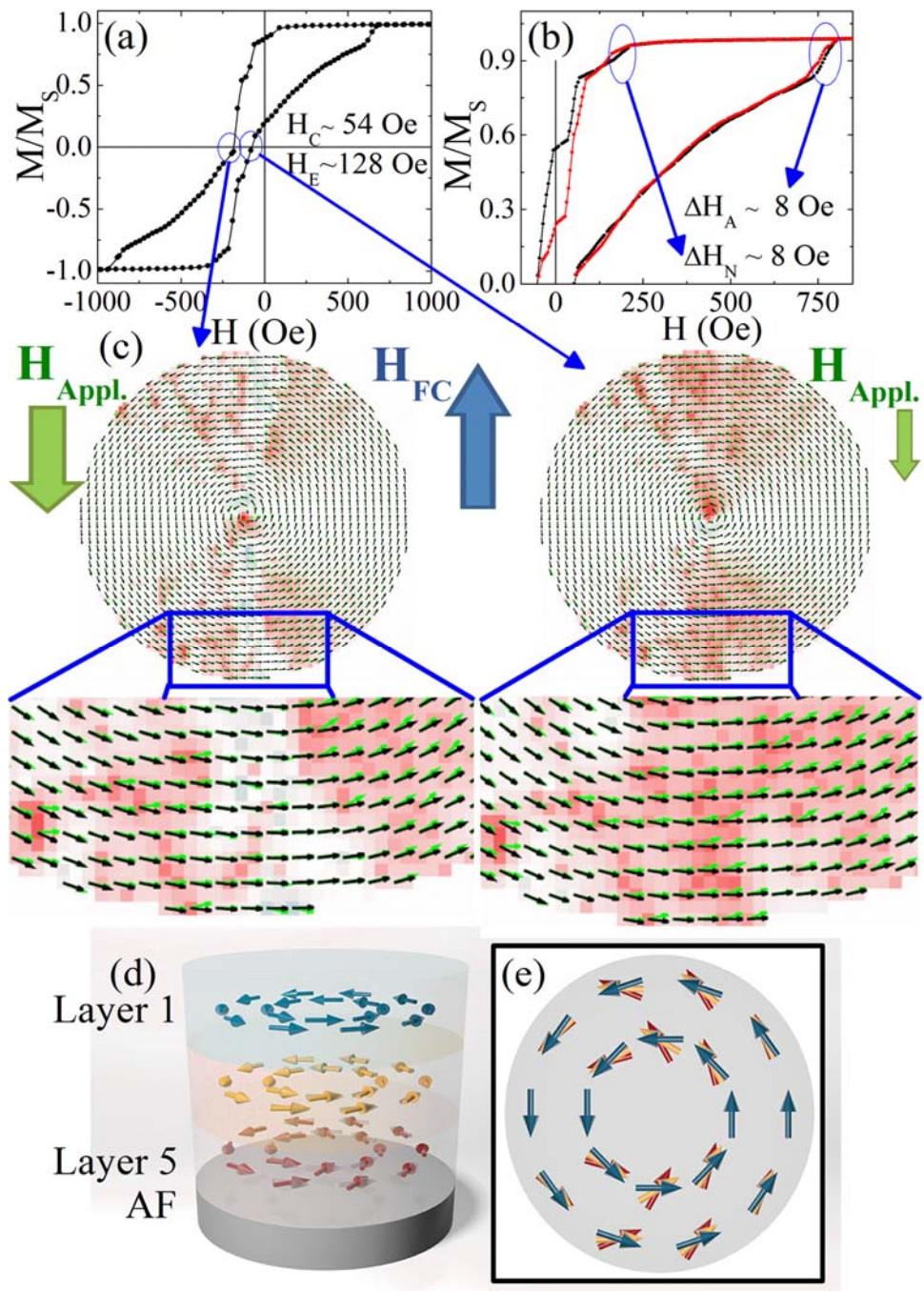

**Figure 3**

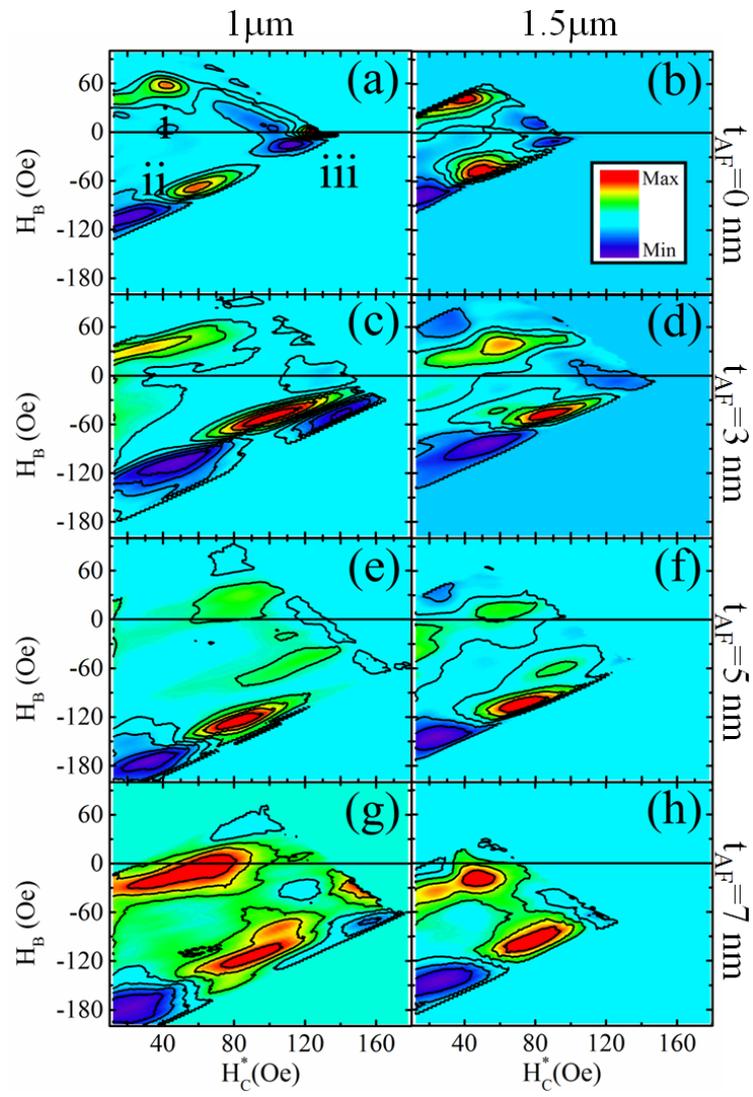

**Figure 4**

# Electronic Supplementary Information

# New Reversal Mode in Exchange Coupled Antiferromagnetic/Ferromagnetic Disks: Distorted Viscous Vortex

Dustin A. Gilbert,[a] Li Ye,[a] Aïda Varea,[b] Sebastià Agramunt-Puig,[c] Nuria del Valle,[c] Carles Navau,[c] José Francisco López-Barbera,[c, d] Kristen S. Buchanan,[e] Axel Hoffmann,[f] Alvar Sánchez,[c] Jordi Sort,[c, g] Kai Liu,[a, *] and Josep Nogués[c, d, g,*]

[a]Physics Department, University of California, Davis, CA, USA

[b]MinD-in2UB, Electronics Department, Universitat de Barcelona, Martí i Franquès 1, 08028, Barcelona, Spain

[c]Departament de Física, Universitat Autònoma de Barcelona, 08193 Bellaterra (Barcelona), Spain

[d]ICN2 – InstitutCatala de Nanociencia i Nanotecnologia, Campus UAB, 08193 Bellaterra (Barcelona), Spain

[e]Department of Physics, Colorado State University, Fort Collins, CO, USA

[f]Materials Science Division, Argonne National Laboratory, Argonne, IL, USA

[g]Institució Catalana de Recerca i Estudis Avançats (ICREA), Barcelona, Spain

*Corresponding authors: kailiu@ucdavis.edu, Josep.Nogues@uab.cat.

**Details of the first order reversal curve (FORC) measurements:**

First, the sample is saturated in a positive field. Then, the field is reduced to a scheduled reversal field, $H_R$, and the magnetization is recorded as the applied field, $H$, is swept back to positive saturation, hence tracing out a single FORC. This sequence is repeated for decreasing values of the reversal field until negative saturation is reached, measuring a family of FORCs where the magnetization, $M$, is recorded as a function of both $H$ and $H_R$. The FORC distribution, $\rho(H,H_R)$ is then calculated by applying a mixed second order derivative:

$$\rho(H, H_R) = -\frac{1}{2M_S}\frac{\partial^2 M(H,H_R)}{\partial H_R \partial H}, \qquad (1)$$

where $M_S$ is the saturation magnetization. The resulting distribution is only non-zero for irreversible switching processes. Recognizing that sweeping $H$ probes the up-switching events and stepping $H_R$ probes the down-switching events, new coordinates can be defined: the local bias $H_B=(H+H_R)/2$ and local coercivity $H_C^*=(H-H_R)/2$.

**Anatomy of the FORC features for vortex, biased vortex and tilted-vortex reversals:**

A schematic of a standard vortex FORC is shown in Fig. S1(a). For a magnetic vortex in a symmetric structure, such as a circular dot, the three main features correspond to (i) nucleation from positive saturation (at $H_R^{(i)}$) and annihilation to positive saturation (at $H^{(i)}$), (ii) annihilation to negative saturation (at $H_R^{(ii)}$) and nucleation from negative saturation (at $H^{(ii)}$), and (iii) subsequent annihilation to positive saturation (at $H_R^{(iii)}=H_R^{(ii)}$, $H^{(iii)}=H^{(i)}$). Feature (ii) is accompanied by a negative feature, extending in the $-H$ direction relative to the positive feature. In the vortex state, the magnetization varies continuously in response to the magnetic field ($dM/dH\neq 0$), whereas in the saturated state the magnetization remains constant ($dM/dH=0$). These 'unmatched' $dM/dH$ slopes [30] leads to a negative feature. The different reversal modes can be distinguished as follows.

*Non-biased vortex*: the nucleation and annihilation fields will be symmetric for positive and negative saturation, thus $H_R^{(i)}=-H^{(ii)}$, $H^{(i)}=-H_R^{(ii)}$ and $H^{(iii)}=H^{(i)}$. Transforming these into ($H_C$, $H_B$) coordinates as described above, $H_C^{(i)}=H_C^{(ii)}$, and $H_B^{(i)}=-H_B^{(ii)}$; feature (iii) occurs at $H_B=0$ at $H_C=H_A$ [Fig. S1(a)].

*Uniformly exchange biased vortex:* as shown in Fig. S1(b), the coordinates of the FORC features can be calculated to be $H_C^{(i)}=H_C^{(ii)}$, and $H_B^{(i)} + H_B^{(ii)}=2H_E$. The FORC features retain their relative arrangement, just offset along the $H_B$ axis by $H_E$.

*Tilted vortex:* as shown in Fig. S1(c), the nucleation fields are equally biased, but not the annihilation fields. Thus $H_R^{(i)}+H_E=-H^{(ii)}+H_E$ and $H^{(i)}+H_E \neq -H_R^{(ii)}+H_E$, and features (i) and (ii) will no-longer remain aligned. Further, feature (iii) will still be intimately coupled to both features, manifesting at $(H_R^{(ii)}, H^{(i)})$.

*Distorted viscous vortex reversal*: as shown in Fig. S1(d) the nucleation and annihilation events both can change and thus (i) and (ii) will not be aligned. However, since the exchange field is very small, the entire FORC distribution is still centered around $H_B\sim 0$.

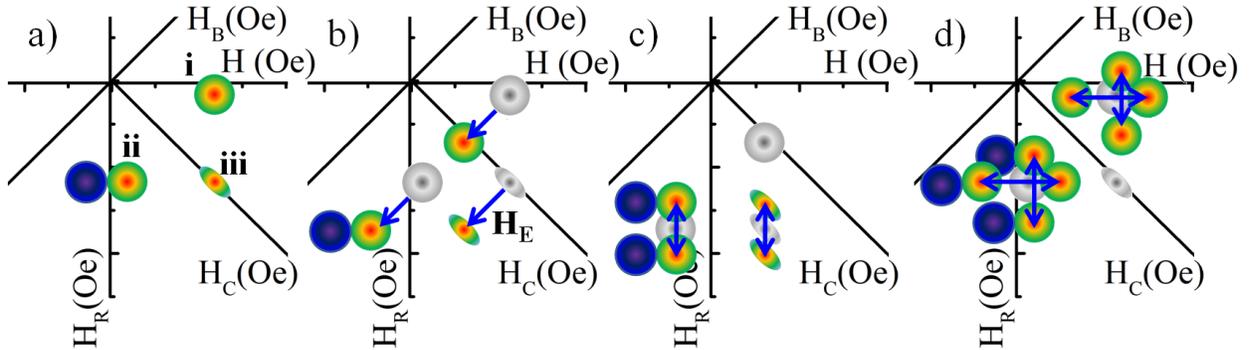

**Fig. S1.** Schematic FORC diagrams for (a) unbiased vortex, (b) biased vortex, (c) tilted vortex, and (d) distorted viscous vortex. Arrows indicate the shift of the major FORC features due to changes in the nucleation and annihilation fields.